\documentclass[useAMS,usenatbib,usegraphicx]{mn2e}

\title{Constraints on dark matter and the shape of the Milky Way dark halo from the 511 keV line}
\author[Y.~Ascasibar et al.]
{Y.~Ascasibar$^1$ \thanks{E-mail: yago@head.cfa.harvard.edu}, P. Jean$^{2,3}$, C.~B\oe hm$^{4,5}$ and J. Kn\"odlseder$^{2,3}$ \\
$^1$Harvard-Smithsonian Center for Astrophysics, 60 Garden St., Cambridge, MA 02138, USA\\
$^2$Centre d'\'Etude Spatiale des Rayonnements, 9 avenue Colonel-Roche, BP 4346, 31028 Toulouse Cedex 4, France\\
$^3$ Universit\'e Paul Sabatier, Toulouse 3, 118, Route de Narbonne, F-31062 Toulouse, France\\
$^4$On leave LAPTH, UMR 5108, 9 chemin de Bellevue, BP 110, 74941 Annecy-Le-Vieux, France\\
 $^5$Physics department, Theory Division, CERN, CH-1211 Geneve 23, Switzerland
}
\newcommand{\be}{\begin{equation}}
\newcommand{\ee}{\end{equation}}
\newcommand{\msun}{{\rm M}_\odot}

\newcommand{\vv}[1]{\bmath{#1}}

\newcommand{\dd}{{\rm d}}

\newcommand{\vir}{_{\rm vir}}
\newcommand{\MLR}{{\rm MLR}}

\newcommand{\sw}{\sin \theta_W}

\newcommand{\dm}{_{\rm dm}}
\newcommand{\dms}{_{\rm dm^\star}}
\newcommand{\mdm}{m\dm}
\newcommand{\mmev}{m_{\rm MeV}}

\newcommand{\fcen}{\Phi_{\rm cen}}
\newcommand{\ftot}{\Phi_{\rm tot}}

\begin{document}

\maketitle

\begin{abstract}
About one year ago, it was speculated that decaying or annihilating
Light Dark Matter (LDM) particles could explain the flux and
extension of the 511 keV line emission in the galactic centre. Here
we present a thorough comparison between theoretical expectations of
the galactic positron distribution within the LDM scenario and
observational data from INTEGRAL/SPI. Unlike previous analyses,
there is now enough statistical evidence to put tight constraints on
the shape of the dark matter halo of our galaxy, if the galactic
positrons originate from dark matter. For annihilating candidates,
the best fit to the observed 511 keV emission is provided by a
radial density profile with inner logarithmic slope
$\gamma=1.03\pm0.04$. In contrast, decaying dark matter requires a
much steeper density profile, $\gamma>1.5$, rather disfavoured by
both observations and numerical simulations. Within the annihilating
LDM scenario, a velocity-independent cross-section would be
consistent with the observational data while a cross-section purely
proportional to $v^2$ can be rejected at a high confidence level.
Assuming the most simplistic model where the galactic positrons are
produced as primaries, we show that the LDM candidate should be
a scalar rather than a spin-1/2 particle and obtain a very stringent
constraint on the value of the positron production cross-section to explain the 511 keV
emission. One consequence is that the value of the fine
structure constant $\alpha$ should differ from that recommended in the
CODATA. This is a very strong test for the LDM scenario and an
additional motivation in favour of experiments measuring $\alpha$
directly. Our results finally indicate that an accurate
measurement of the shape of the dark halo profile could have a
tremendous impact on the determination of the origin of the 511 keV
line and vice versa.
\end{abstract}


%--------------------------------------------------------------------------
  \section{Introduction} \label{secIntro}
%--------------------------------------------------------------------------

An emission line at 511 keV was detected at the galactic centre
three decades ago \citep{Johnson72}. Its identification as an
electron-positron annihilation line followed as soon as
high-resolution spectrometers became available \citep{Leventhal78},
but the origin of low-energy galactic positrons is still a matter of
heated debate. The latest observations of the annihilation emission
have been performed by the SPI spectrometer aboard the
INTEGRAL\footnote{INTEGRAL (International Gamma Ray Laboratory) is
an ESA's gamma ray observatory launched in October 2002.} satellite.
A total flux of $\approx10^{-3}$ photons s$^{-1}$ cm$^{-2}$ was
measured, in agreement with previous estimates. The morphology of
the galactic bulge emission could be fit by a Gaussian with $\sim$10
\degr FWHM. A disc component was recently detected by INTEGRAL/SPI
(Kn\"odlseder et al. 2005) but this emission can be attributed to
the $\beta^+$ decay of the radioactive species $^{26}$Al and
$^{44}$Ti, which are produced by massive stars in the disc.

Several astrophysical sources have been proposed in the literature
to explain the low-energy positrons from the bulge, such as radioactive nuclei
expelled by stars (supernovae, hypernovae, novae, Wolf-Rayet stars
and red giants) and collapsed objects (neutron stars or black
holes). Nevertheless, most of these sources \citep[see][ and
references therein]{Knodlseder05} cannot account for the observed
morphology, due to the large bulge-to-disc ratio of the emission,
which suggests an old stellar population origin, unless rather
elaborate mechanisms (e.g. jets, propagation) are invoked.

On the other hand, the presence of low-energy positrons could be
explained by Dark Matter (DM) annihilations \citep{boehmhooper} or
decays \citep{hooper,pospelov}. The present paper focuses on such
scenarios, which require light dark matter particles (i.e. with a
mass $\mdm \la 100$ MeV, depending on their exact nature) in order
to reproduce the observational data.

The smallness of the DM mass might appear surprising to many.
Indeed, not so long ago, most of the community thought that
annihilating DM particles should be heavier than a few GeV because
of the Lee-Weinberg limit \citep{leew}, which states that if DM is a
stable fermion coupled to heavy particles (such as the Z and W gauge
bosons) then its mass should exceed that of the proton; otherwise it
would overclose the universe. There are possible ways to evade the
Lee-Weinberg limit though but the fact that theoretically motivated
DM candidates, such as the lightest neutralino, were naturally very
heavy did not encourage the community to investigate the lighter
range.

The window for Light Dark Matter (LDM) particles suddenly opened
when it was realized that scalar candidates with a mass  from a few
MeV to a few GeV, coupled to heavy fermions ($F$) or to light
neutral particles (neutral gauge bosons $Z'$, somewhat analogous to
the $Z$ gauge boson), could also yield the observed relic density.
However, the introduction of LDM particles immediately faces an
embarassing problem:  their annihilations are expected to produce
too many low-energy gamma rays in our galaxy, compared to what has
been observed. To be on the safe side, the present-day annihilation
cross-section must be about five orders of magnitude (times
$\mmev^2$, where $\mmev\equiv\mdm c^2/\rm{1~MeV}$) smaller than it
was in the primordial universe \citep{bens}.

Such a condition can be easily satisfied if the channel associated
to the exchange of heavy fermions is suppressed with respect to the
one due to the new gauge boson. Indeed, the annihilation
cross-section associated with a $Z'$ exchange is proportional to the
square of the DM velocity, which -- in the Milky Way -- is at least
two or three orders of magnitude smaller than in the primordial
universe (i.e. before DM became non-relativistic). Hence a
velocity-dependent cross-section can satisfy both the relic density
criterion and the gamma ray constraint. In contrast, the
cross-section arising from heavy-fermion exchange does not depend at
all on velocity; it remains constant at any epoch. The LDM scenario
is thus viable if -- as initially proposed -- the  $F$ exchange is
at least five orders of magnitude (times $\mmev^2$) smaller than the
$Z'$ cross-section at early times. Note, though, that the
contribution of the fermion exchange to the total annihilation
cross-section could become dominant as velocities become
non-relativistic.

Although the parameter space allowed for LDM is still quite broad,
it is encouraging to find out that this model, originally built to
be invisible (more precisely, to escape the low-energy gamma ray
constraint) is able to explain the observed properties of the 511
keV line without adding any new components nor changing the mass
range initially proposed. Moreover, the existence of LDM particles
might also explain several particle physics measurements \citep[see
e.g.][]{boehmyago}. The detailed analysis of the 511 keV line
presented here provides new -- and relatively tight -- constraints
on some of the LDM parameters, greatly enhancing the predictive
capability of the model in other fields.

Conversely, independent confirmation of the LDM scenario from
particle physics experiments would have important astrophysical
consequences. In fact, if dark matter annihilation (light or
otherwise) turns out to be the main source of galactic low-energy
positrons, the observed morphology of the 511 keV emission line
would constitute an excellent probe of the shape of the Milky Way
dark matter halo, whatever the exact nature of dark matter particles
might be. The present study shows that several robust conclusions
can already be derived from current INTEGRAL/SPI data.

Previous analyses \citep[e.g.][]{Jean03} have shown that a point
source can be ruled out a high confidence level. If positrons cannot
travel a long distance before annihilating, that would seem to
suggest that the galactic halo cannot be too `cuspy'. On the other
hand, a very flat profile would not match the observed morphology of
the emission either, because its FWHM tells us that most of the
positrons are generated within 1 kpc from the galactic centre.
According to \citet{boehmyago}, the observed morphology is well
described if the dark matter halo of the Milky Way follows a
\citet{NFW97} profile. In this paper, we will attempt to constrain
both the nature of dark matter and its distribution within the Milky
Way halo as independently as possible.

Section~\ref{secDM} focuses on the main features of our dark matter
model. The description of the Milky Way halo is discussed in
Section~\ref{secMW}. Section~\ref{secObs} is devoted to the
comparison between the theoretical positron distribution and the
observed flux. The results of our likelihood analysis are reported
in Section~\ref{secResults}, while  Section~\ref{secDiscus}
discusses additional issues, indirectly related to the present work.
Finally, our main conclusions are briefly summarized in
Section~\ref{secConclus}.

%--------------------------------------------------------------------------
  \section{Dark matter characteristics} \label{secDM}
%--------------------------------------------------------------------------

\begin{figure}
\centering
\includegraphics[width=8cm]{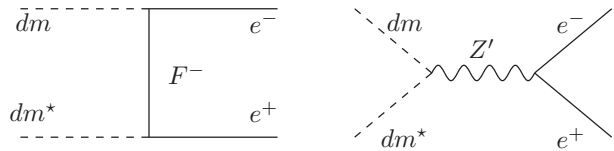}
\caption{Dark matter annihilation through the exchange of a charged heavy fermion $F$ (left) or a neutral light gauge boson $Z'$ (right).}
\label{figFey}
\end{figure}

Our study is based on the assumption that most galactic positrons
originate from the decays or annihilations of LDM particles. The
number density of positrons produced per unit of time is then
dictated by the number density of dark matter particles, $n\dm$,
times their decay/annihilation rate into a pair $e^+ e^-$. The
latter are given by  $\Gamma_{\rm d}$ or $\Gamma_{\rm a} = \langle
\sigma v_r\rangle n\dms$ respectively, where $\langle \sigma
v_r\rangle$ is the thermal average of the annihilation cross-section
times the DM relative velocity.
The two annihilation channels mentioned in the introduction are represented in Figure~\ref{figFey}.

Since the energy of the particles in the final state is imposed by
kinematics, the only quantities that may appear in the formula of
the annihilation cross-section are the masses of all particles
involved, their couplings and the energy $E\dm$ of the initial
state.
Since DM particles are nowadays non-relativistic, $E\dm\approx \mdm\,c^2 + \mdm\,v^2/2$, so the average cross-section can be re-written as $\langle\sigma v_r\rangle \approx a+bv^2+O(v^4)$, where both $v$ and $v_r$ are expressed in units of the speed of light, $c$.

In our case, the $F$-exchange cross-section has both an $a$- and a
$b$-term; both of the same order of magnitude. In contrast, the $Z'$
exchange gives rise to a pure velocity-dependent cross-section (i.e.
$a=0$). A combination of both ingredients ($Z'$ and heavy fermions)
thus provides a total annihilation cross-section with $a\neq b\neq
0$. Moreover, the cross-section through the $Z'$ exchange depends on
$\mdm$, while the cross-section associated with the exchange of
heavy particles is proportional to $1/m_{F}^2$ or $\mdm^2/m_F^4$,
depending on whether DM is a spin-0 or spin-1/2 particle,
respectively \citep{bf}.

There are two independent constraints on the values of $a$ and $b$ (or more precisely on $\langle \sigma v_r \rangle$).
On one hand, the relic density criterion imposes that the annihilation cross-section at the time of the chemical decoupling (i.e. when $T\sim\mdm$ and $v\dm\sim c$) is about $\langle \sigma v_r \rangle\sim 10^{-26}$~cm$^3$~s$^{-1}$ so that $\Omega\dm h^2 \sim 0.1$ nowadays.
On the other hand, the amount of low-energy gamma rays produced at the present day in the centre of the Milky Way should not exceed the values observed by COMPTEL and EGRET \citep{Strong00}.
An estimate of the gamma ray flux generated by LDM annihilations has been computed by \citet{bens} in terms of $\langle\sigma v_r\rangle$, yielding\footnote{The flux obtained by \citet{bens} was actually overestimated by at least a factor four, since it was assumed that the number densities of dark matter particles and anti-particles were given by $n\dm = n\dms =\rho\dm/\mdm$ instead of $n\dm = n\dms = \rho\dm/(2 \mdm)$.} an upper limit
$\langle\sigma v_r\rangle \la 10^{-31} \mmev^2\ {\rm cm^3\ s^{-1}}$.
Combining both constraints, we obtain
\be
\langle \sigma v_r \rangle_{\rm prim} \approx
     a + b/9 \sim 10^{-26} \ \mbox{cm}^3 \mbox{s}^{-1}
\label{eqRelic}
\ee
in the primordial universe (with $v\dm\approx c/3$) and
\be
\langle \sigma v_r \rangle_{\rm now} \approx
     a + v_0^2\,b \la 10^{-31} \mmev^2\ \mbox{cm}^3 \mbox{s}^{-1}
\label{eqGamma}
\ee
in our Galaxy, where $v_0 \sim 10^{-3}c$ and assuming that every DM annihilation into electron-positron instantaneously produces two photons with the maximal energy.
Given that the photon production is expected to be
through the final state radiation (i.e. a radiative correction to
the annihilation process), we are clearly overestimating the gamma
ray flux, making the constraint more stringent than it should be
(see the discussion in Section~\ref{secGamma}).

In any case, it seems clear that, for $\mdm\la 100$~MeV, the annihilation cross-section at the present time must be suppressed with respect to its primordial value in order to satisfy both the relic density (\ref{eqRelic}) and the gamma ray (\ref{eqGamma}) conditions, which lead lead to $a<10^{-31}\,\mmev^2\ {\rm cm^3\ s^{-1}}$ and $b\sim10^{-25}\ {\rm cm^3\ s^{-1}}$.
A velocity-dependent cross-section is thus necessary for LDM candidates below $\sim100$~MeV.

The gamma-ray constraint depends nevertheless on the shape of the dark halo profile.
In particular, equation (\ref{eqGamma}) above was derived from a NFW profile.
Taking a flatter halo
lessens these constraints and one finds that a velocity-independent
cross-section can match both the relic density and the gamma ray
criteria if $\mdm> 20$~MeV \citep{komatsu}. Hence, if the dark halo radial densiy profile of the
Milky Way turns out to be flat, it would be possible to set $b=0$
and get rid of the $Z'$.

In the following, the constants $a$ and $b$ will be normalized to
$10^{-26}\ {\rm cm^3\ s^{-1}}$, yielding the notation $a_{26}$ and
$b_{26}$, respectively. For decaying LDM particles, the decay rate
$\Gamma_{\rm d}$ will also be normalized to $10^{-26}\ {\rm s^{-1}}$
\citep{hooper,pospelov} yielding the notation $\Gamma_{26}$.

%--------------------------------------------------------------------------
  \section{The Milky Way dark halo} \label{secMW}
%--------------------------------------------------------------------------

Pioneering analytical studies based on the spherical collapse
formalism \citep{GG72,Gunn77} predicted that dark matter haloes
ought to be described by a single power-law density profile,
$\rho(r) \propto r^{-\gamma}$, with $\gamma$ ranging from 2 to 2.25
\citep[e.g.][]{FG84,Bertschinger85}. Such result appeared quite
compelling at the time, as it could straightforwardly explain the
flatness of the rotation curves observed in spiral galaxies.

Later work based on numerical N-body simulations showed that the
density profile was shallower than isothermal ($\gamma<2$) as
$r\to0$, and steeper ($\gamma\approx3$) as $r\to\infty$. Typical
values measured for the asymptotic logarithmic slope at the centre
range from $\gamma=1$ \citep[][hereafter NFW]{NFW97} to $\gamma=1.5$
\citep[][hereafter M99]{Moore99}. The very existence of an
asymptotic behaviour has recently been questioned by several studies
\citep[e.g.][]{Power03,Hayashi04,Navarro04}, in which the density
profiles found in high-resolution simulations are reported to become
progressively shallower inwards.

Despite the significant uncertainty on the shape of the density
profile near the centre\footnote{Note that, in this region,
numerical experiments are severely hampered by two-body relaxation
\citep[see e.g.][]{Diemand04} and discreteness effects
\citep{Binney04}.}, there is general agreement in that the dark
matter distribution within a spherically symmetric halo can be well
fitted by a `universal' function with a small number of free
parameters, and that the same functional form is valid for a broad
range of halo masses and underlying cosmologies. Most of the
analytical formulae proposed in the literature can be cast in the
form \be \rho(r)= \frac{ \rho_0 }{
{(r/r_0)}^{\gamma}\,{\left[1+{(r/r_0)}^\alpha\right]}^{(\beta-\gamma)/\alpha}
}, \label{eqrho} \ee where $\rho_0$ and $r_0$ are a characteristic
density and radius of the halo, $\gamma$ is the asymptotic
logarithmic slope at the centre, $\beta$ is the slope as
$r\to\infty$ and $\alpha$ controls the exact shape of the profile in
the intermediate regions around $r_0$.

Many different sets of values have been suggested for these
parameters. The most notable discrepancies concern the value of
$\gamma$. In particular, observed rotation curves of dwarf spiral
and low surface brightness galaxies tend to favour flat profiles
($\gamma\approx0$), which has been often signalled as a genuine
crisis of the CDM scenario \citep[e.g.][]{FloresPrimack94,Moore94}.
Recent analyses show that observational data may actually be
consistent with steeper profiles once the effects of inclination,
non-circular orbits and triaxiality of the dark matter haloes are
accounted for \citep{Hayashi04,Hayashi_05}, but the controversy is
still unresolved \citep[e.g.][]{Gentile04,deblok}.

In the Milky Way, it has been argued \citep[][hereafter
BE]{BinneyEvans01} that the microlensing optical depth towards the
galactic centre reported by the MACHO collaboration \citep{Alcock00}
would be incompatible with a $\gamma\ga0.3$ dark matter halo. A
recent revision of the MACHO results yields a lower optical depth,
in somewhat better agreement with the values found by other
experiments \citep[see e.g.][for a recent review]{Sumi_05}. Although
a low optical depth could be consistent with steeper profiles, it is still
unclear whether it would be compatible with the asymptotic slopes
characteristic of numerical haloes \citep[see e.g.][]{Binney_04_IAU}.

On the other hand, the existence of a black hole at the centre of
our galaxy \citep{joe} and the adiabatic contraction of the dark
matter component due to the presence of baryons \citep{Blumenthal86}
are expected to increase the central dark matter density, leading to
a very steep profile in the innermost regions.

\begin{table}
\caption{
Radial density profiles of the Milky Way dark halo considered in the present work, according to the parametrization given by expression (\ref{eqrho}).
}
\label{tabProf}
\centerline{
\begin{tabular}{cccccc}
\hline
 MW & $\alpha$ & $\beta$ & $\gamma$ & $r_0$ [kpc] & $\rho_0$ [GeV cm$^{-3}$] \\
\hline
 ISO & 2   & 2 &  0   &  4.0 & 1.655 \\
 BE  & 1   & 3 & 0.3  & 10.2 & 1.459 \\
 NFW & 1   & 3 &  1   & 16.7 & 0.347 \\
 M99 & 1.5 & 3 & 1.5  & 29.5 & 0.0536 \\
\hline
\end{tabular}
}
\end{table}

\begin{figure}
\centering
\includegraphics[width=8cm]{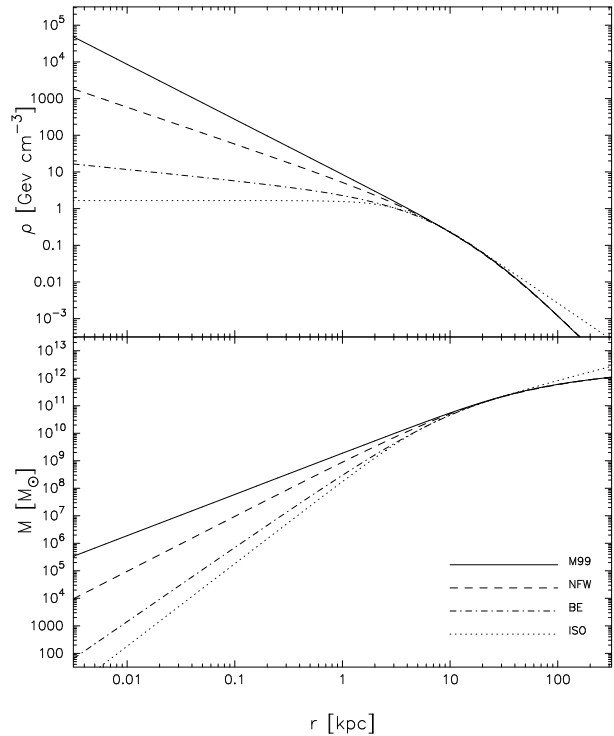}
\caption{ Top panel: Densty profiles considered in the present work.
All of them are described by equation (\ref{eqrho}), with values of
the parameters given in Table~\ref{tabProf}. Bottom panel: The
corresponding cumulative mass profiles. } \label{figProf}
\end{figure}

Most theoretical predictions of the gamma ray emission due to dark matter, including those performed in a supersymmetric framework (i.e. for heavy DM candidates), are based on `cuspy' density profiles, since these maximize the expected flux.
Given the present uncertainties, though, we have followed a completely different approach, trying to constrain the shape of the Milky Way dark matter halo as independently as possible from the precise nature of dark matter particles.

We therefore have considered four different models of the density profile of our galaxy, each one featuring a different asymptotic slope at the centre: in order of decreasing `cuspiness', M99, NFW, BE and a non-singular isothermal sphere (hereafter ISO).
Their corresponding parameters are summarized in Table~\ref{tabProf}, and the density and cumulative mass profiles are depicted in Figure~\ref{figProf}.
In addition, we also consider a family of models in which $\alpha$ and $\beta$ are fixed to 1 and 3, respectively, while $\gamma$ is varied in uniform steps $\Delta\gamma=0.05$.

In all cases, the normalization of the models, $\rho_0$, is set by
imposing a local dark matter density
$\rho(r_\odot)=0.3$~GeV~cm$^{-3}$, with $r_\odot=8.5$~kpc. The
characteristic radius $r_0$ has been chosen so that the virial
radius and mass are $R\vir\approx260~{\rm kpc}$ and $M\vir\approx
10^{12}~\msun$. Note that the ISO model can only approximately
satisfy this condition.

The characteristic velocity of dark matter particles is also a
necessary ingredient in our model of the Milky Way, as the LDM
annihilation cross-section associated to $Z'$ exchange explicitly
depends on this quantity. In many studies, the rough estimate
$v\dm\sim 10^{-3}c$ is assumed to be accurate enough, but this is
definitely not true for our present analysis. The positron emission
arising from the velocity-dependent term of the cross-section will
be sensitive to the product $\rho^2(r)\sigma^2(r)$. Since we are
trying to constrain the shape of the density profile, it is
extremely important to properly account for the radial variation of
the velocity dispersion.

\begin{figure}
\centering
\includegraphics[width=8cm]{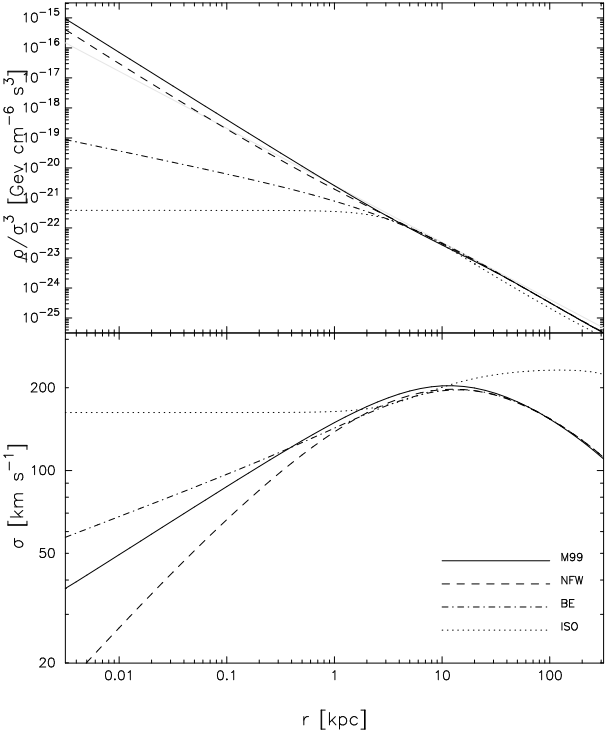}
\caption{ Velocity dispersion profile obtained from the Jeans
equation for each of the models in Table~\ref{tabProf}. Grey line on
the top panel shows expression (\ref{eqSigma}). } \label{figSigma}
\end{figure}

Indeed, the velocity dispersion profile of dark matter haloes is
known to change systematically with radius. As shown by
\citet{TN01}, the phase-space density profile found in N-body
simulations follows an approximate power law over several orders of
magnitude in $r$. This result has been confirmed by \citet{Rasia04}
and \citet{Ascasibar04} for haloes of very different mass. Following
the latter, the density and velocity dispersion profiles would be
subject to the phenomenological relationship

\be
\frac{\rho(r)}{\sigma^3(r)} = 10^{1.46\pm0.04}
\frac{\rho_c}{V\vir^3} \left( \frac{r}{R\vir} \right)^{-1.90\pm0.05}
\ee

where $R\vir$ is the virial radius of the halo and we define
$V\vir^2\equiv GM\vir/R\vir$. The cosmology-related quantities
$\rho_c=1.36\times10^{11}~\msun$~Mpc$^{-3}$ and $M\vir\approx
\frac{4\pi}{3}100\rho_cR\vir^3$ are the critical density and virial
mass corresponding to a `concordance' $\Lambda$CDM universe
($\Omega_{\rm m}=0.3$, $\Omega_\Lambda=0.7$, $h=0.7$). Substituting
$R\vir=260$ kpc for the Milky Way, we obtain
$M\vir\approx10^{12}~\msun$ and $V\vir\approx130$~km~s$^{-1}$. The
velocity dispersion profile of our galaxy is thus expected to vary
with radius as \be \frac{ \sigma^2(r) }{c^2 } \approx 6\times10^{-8}
{\left[\,
  \frac{\rho(r)}{1~{\rm GeV~cm^{-3}}}
  \left(\frac{r}{1~{\rm kpc}}\right)^{1.9}
\,\right]}^{2/3}.
\label{eqSigma}
\ee

This formula has been used by \citet{boehmyago} to estimate the
characteristic velocity of dark matter particles. The radial
dependence of the velocity dispersion significantly reduces the
emission from the boson-exchange channel, being roughly equivalent
to an effective flattening of the density profile.

However, expression (\ref{eqSigma}) is a mere fit to the
coarse-grained phase-space density profiles found in numerical
simulations, and thus its validity has only been tested for `cuspy'
density profiles. A more self-consistent approach is followed in the
present study, where we derive the velocity dispersion profiles from
the spherically-symmetric Jeans equation, \be
\frac{1}{\rho(r)}\frac{\dd[\rho(r)\,\langle v_{\rm ra}^2(r)
\rangle]}{\dd r} +2\beta(r) \frac{\langle v_{\rm ra}^2(r)
\rangle}{r} =-\frac{GM(r)}{r^2}, \ee where $v_{\rm ra}$ is the
radial component of the velocity and the anisotropy parameter
$\beta(r)$ is defined as $\beta(r)\equiv1-\langle v_{\theta}^2(r)
\rangle/\langle v_{\rm ra}^2(r) \rangle$. Assuming no radial infall,
an isotropic velocity ellipsoid and vanishing velocity dispersion at
infinity, \be
\sigma^2(r)=\frac{3}{\rho(r)}\int_r^\infty{\rho(r)\frac{GM(r)}{r^2}}.
\ee

The corresponding profiles are plotted in Figure~\ref{figSigma}. As
can be readily seen, equation (\ref{eqSigma}) provides a fair
approximation for NFW and M99 models, but it is certainly not
adequate for shallower density profiles.

%--------------------------------------------------------------------------
  \section{Comparison with SPI data} \label{secObs}
%--------------------------------------------------------------------------

According to the model outlined so far, the rate at which new
positrons are created is \be \dot n_{\rm e^+}=n\dm\,\Gamma \ee where
$\Gamma$ corresponds to the decay rate $\Gamma_{\rm d}$ for decaying
dark matter and to the annihilation rate $\Gamma_{\rm a}=(a+b\,v\dm^2) n\dms$ for annihilating LDM particles, with $n\dms$ being the number density of dark matter anti-particles.

These positrons will be relativistic at the moment of their creation
($E_{\rm e^+}\sim\mdm c^2$). However, they can efficiently lose
their energy through collisional ionization or excitation in neutral
Hydrogen and by interaction with plasma waves in ionized
interstellar medium. We will make the approximation that they can
only travel a short distance before becoming non-relativistic and
annihilate with an electron at rest. Such an assumption is not very
realistic for the outer regions of the galaxy, but it is perfectly
reasonable for the galactic bulge, where most of the observed
emission comes from.

Both OSSE \citep{Kinzer01}, TGRS \citep{Harris98} and SPI
\citep{Churazov05} measurements indicate that approximately 93 per
cent of the positrons annihilate through positronium formation. In
this channel, $3/4$ of the annihilations take place in the
orthopositronium state, yielding 3 photons with $E<511$ keV each,
while the remaining $1/4$ annihilate in the parapositronium state,
producing 2 photons with $E=511$ keV. The remaining 7 per cent that
do not form a positronium annihilate directly into 2 photons with
$E=511$ keV. Consequently, the total number of 511 keV photons
produced per unit time would be given by \be \dot n_\gamma=
2\,(0.07+0.93/4)\,\dot n_{\rm e^+} = 0.605\,\dot n_{\rm e^+}. \ee

The predicted intensity distribution for any particular model of the
Milky Way dark halo can be thus computed as the integral along the
line of sight, as a function of galactic longitude $l$ and latitude
$b$, of the emissivity $\dot n_\gamma(\vv{r})$, \be I(l,b)=
\frac{1}{4 \pi}\int_0^\infty\!\dot n_\gamma(\vv{r})\,\dd s, \ee
where the spatial dependence arises through the radial density and
relative velocity profiles
$n\dm(r)=n\dms(r)=\rho\dm(r)/(2\mdm)$ and
$v\dm^2(r)\approx\sigma^2(r)$. The total photon flux at the earth is
simply \be \Phi=\int I(l,b)\,\dd\Omega. \ee

Although the resulting sky map will obviously preserve the spherical
symmetry of our models, the morphology of the emission (more
specifically, its concentration) depends on the shape of the Milky
Way halo as well as on the DM annihilation cross-section or decay
rate\footnote{Note that the precise distribution of baryons and
their chemical composition should also be included in a realistic
model of positron propagation.}. Observations are mostly sensitive
to the details of the central part, where the intensity of the 511
keV line is highest. Dark matter in the outermost regions can make a
significant contribution to the total flux, but the intensity of the
emission is so low that it is difficult to discriminate from the
instrumental background.

\begin{table}
\caption{
Theoretical photon fluxes (in ${\rm cm^{-2}~s^{-1}}$) expected for different halo profiles and DM type.
} \label{tabfluxes} \centerline{
\begin{tabular}{cccc}
\hline
DM & MW & $\ftot\,\mmev^2$ & $\fcen\,\mmev^2$ \\
\hline
 $\Gamma_d$ & ISO & $0.0459\ \Gamma_{26}$ & $0.0201\ \Gamma_{26}$  \\
          & BE  & $0.0439\ \Gamma_{26}$ & $0.0195\ \Gamma_{26}$ \\
          & NFW & $0.0478\ \Gamma_{26}$ & $0.0232\ \Gamma_{26}$ \\
          & M99 & $0.0512\ \Gamma_{26}$ & $0.0262\ \Gamma_{26}$ \\
 %     $a$ & ISO & $41.1\ a_{26}$                & $25.4 \ a_{26}$ \\
       $a$ & ISO & $5.13\ a_{26}$                & $3.175 \ a_{26}$ \\
%          & BE  & $41.7\ a_{26}$                & $26.6 \ a_{26}$ \\
           & BE  & $5.21\ a_{26}$                & $3.325 \ a_{26}$ \\
%          & NFW & $76.2\ a_{26}$                & $59.9 \ a_{26}$ \\
           & NFW & $9.52\ a_{26}$                & $7.487 \ a_{26}$ \\
%          & M99 & $213 \ a_{26}$                & $196  \ a_{26}$ \\
           & M99 & $26.6 \ a_{26}$                & $24.5 \ a_{26}$ \\
%      $b$ & ISO & $1.54 \times 10^{-5}\ b_{26}$ & $8.96 \times 10^{-6}\ b_{26}$ \\
           $b$ & ISO & $1.92 \times 10^{-6}\ b_{26}$ & $1.12 \times 10^{-6}\ b_{26}$ \\
%          & BE  & $1.50 \times 10^{-5}\ b_{26}$ & $8.91 \times 10^{-6}\ b_{26}$ \\
           & BE  & $1.87 \times 10^{-6}\ b_{26}$ & $1.11 \times 10^{-6}\ b_{26}$ \\
%          & NFW & $2.35 \times 10^{-5}\ b_{26}$ & $1.68 \times 10^{-5}\ b_{26}$ \\
           & NFW & $2.93 \times 10^{-6}\ b_{26}$ & $2.10 \times 10^{-6}\ b_{26}$ \\
%          & M99 & $4.72 \times 10^{-5}\ b_{26}$ & $3.98 \times 10^{-5}\ b_{26}$ \\
           & M99 & $5.90 \times 10^{-6}\ b_{26}$ & $4.97 \times 10^{-6}\ b_{26}$ \\
\hline
\end{tabular}
}
\end{table}

In order to compare with observational data, intensity maps $I(l,b)$ have been computed for $|l|<60\degr$ and $|b|<50\degr$.
The total flux within this area is denoted by $\ftot$.
However, a fairer comparison with the flux measured by the satellite is given by the central $33\degr$ ($\sim 1$ steradian).
We shall quote this flux as $\fcen$.
The values of $\ftot$ and $\fcen$ expected for each combination of dark matter type and radial density profile are given in Table~\ref{tabfluxes}.

Our analysis has been performed on the December 10, 2004 public
INTEGRAL data release, which consists of $\approx$309 days of
observations. In order to reduce systematic uncertainties in the
analysis, we exclude observation periods with strong instrumental
background fluctuations\footnote{These background variations are
generally due to solar flares or exit and entry of the observatory
in radiation belts.}. The total effective exposure time after
cleaning is 15.3 Ms. The exposure is quite uniform in the central
regions of our Galaxy ($|l|<50\degr$ and $ |b|<30\degr$).

We use a maximum likelihood algorithm to compare the theoretical sky maps with the INTEGRAL/SPI data.
This method has already been applied to SPI data to characterize the morphology of the annihilation.
A detailed description can be found in \citet{Knodlseder05}.

Briefly, the normalization of each theoretical model is fitted to
reproduce the measured rate in the $507.5-514.5$~keV energy range,
taking into account an instrumental background model, the pointings
history and the spatial and energy response functions of SPI.
Normalized maps have been convolved with the response function,
providing the expected number of counts in each detector as a
function of the pointing periods. We then find the intensity that
 maximizes the log likelihood. We subtract from this log
likelihood  $L_{1}$ the log likelihood  $L_{0}$ that is calculated
under the hypothesis that there is no 511 keV source. Multiplication
by a factor of 2 provides the maximum log-likelihood ratio, $\MLR =
2 \, (L_{1}-L_{0})$.

\begin{table}
\caption{
Results of the model-fitting analysis.
Fluxes in units of $10^{-3}\ {\rm cm^{-2}~s^{-1}}$.
}
\label{tabSingle}
\centerline{
\begin{tabular}{ccccc}
\hline
DM & MW & $\ftot$  &$\fcen$ & MLR \\
\hline
 $\Gamma_d$ & ISO &  $6.82 \pm 0.58$ & $2.95 \pm 0.25$ & 135.2 \\
          & BE  &  $7.23 \pm 0.57$ & $3.18 \pm 0.25$ & 167.3 \\
          & NFW &  $7.36 \pm 0.46$ & $3.53 \pm 0.22$ & 261.2 \\
          & M99 &  $6.86 \pm 0.37$ & $3.48 \pm 0.19$ & 332.0 \\
      $a$ & ISO &  $5.55 \pm 0.33$ & $3.40 \pm 0.20$ & 282.8 \\
          & BE  &  $4.98 \pm 0.27$ & $3.16 \pm 0.17$ & 353.6 \\
      & NFW &  $2.49 \pm 0.11$ & $1.95 \pm 0.09$ & 459.9 \\
          & M99 &  $0.83 \pm 0.04$ & $0.76 \pm 0.04$ & 339.2 \\
      $b$ & ISO &  $6.00 \pm 0.38$ & $3.46 \pm 0.22$ & 258.3 \\
          & BE  &  $5.76 \pm 0.32$ & $3.40 \pm 0.19$ & 305.7 \\
          & NFW &  $3.61 \pm 0.18$ & $2.57 \pm 0.13$ & 422.4 \\
          & M99 &  $1.57 \pm 0.07$ & $1.32 \pm 0.06$ & 430.0 \\
\hline
\end{tabular}
}
\end{table}

\begin{table}
\caption{
MLR and total flux (normalized to $10^{-3}\ {\rm cm^{-2}~s^{-1}}$) for the combined models $\phi_a I_a + \phi_b I_b$, where $I_a$ and $I_b$ are the intensity distributions associated with $a$ and $b$-terms, respectively.
}
\label{tabCombined}
\centerline{
\begin{tabular}{crrcc}
\hline
DM &$\phi_a$~~\ ~~ & $\phi_b$~~\ ~~  & $\fcen$  & MLR \\
\hline
ISO & $28.07\pm2.68$ & $-26.35\pm2.85$ & 1.72 & 368.1 \\
BE  & $15.11\pm1.32$ & $-13.95\pm1.53$ & 1.16 & 437.1 \\
NFW & $ 2.42\pm0.39$ & $ -0.67\pm0.53$ & 1.75 & 461.4 \\
M99 & $-1.00\pm0.16$ & $  2.82\pm0.25$ & 1.82 & 464.8 \\
\hline
\end{tabular}
}
\end{table}

Results of the model-fitting procedure are presented in
Table~\ref{tabSingle} and Table~\ref{tabCombined}. As in the
theoretical models, $\ftot$ is the total flux of the map, integrated
over the whole solid angle, while $\fcen$ is restricted to an
aperture of $33\degr$. When comparing two models, the one with the
largest MLR can be said to explain the data better than the other,
although differences $\Delta_\MLR<10$ are, in principle, not very
significant.

Strictly speaking, the posterior probability distribution for the flux, given the data and a particular model, is given by:
\be
P(\Phi|D,M)~\dd\Phi= \frac{ P(D|M,\Phi) P(\Phi|M)~\dd\Phi } { P(D|M) },
\ee
where $D$ is the data, $M$ is the model and $\Phi$ is the value of the flux, which can be regarded as a vector in the two-parameter models.
$P(D|M,\Phi)$ is the likelihood of the data given $M$ and $\Phi$, $P(\Phi|M)~\dd\Phi$ is the prior probability distribution of the flux, and  $P(D|M)$ is the Bayesian evidence of model $M$,
\be
P(D|M)= \int P(D|M,\Phi) P(\Phi|M)~\dd\Phi.
\ee

The Bayesian evidence provides a measure of how well the model explains the data.
More precisely, model $M_1$ can be considered $P(D|M_1)/P(D|M_0)$ times more likely than model $M_0$.
Since the likelihood $P(D|M,\Phi)$ is strongly peaked around the best-fitting value of $\Phi$, the evidence is not very sensitive to the precise shape assumed for the prior.
Morevoer, for practical purposes we can approximate $P(\Phi|D,M)$ by a Gaussian centered around the best-fitting $\Phi$ with $\sigma\approx\Delta\Phi$, so that the evidence can be computed as
\be
P(D|M)\propto e^{\rm MLR}~\sqrt{2\pi}\Delta\Phi
\ee
for the models with one parameter, and
\be
P(D|M)\propto e^{\rm MLR}~2\pi\Delta\phi_a\Delta\phi_b
\ee
for the composite models in Table~\ref{tabCombined}.

It is evident from the figures in Table~\ref{tabSingle} that we can neglect the effect of $\Delta\Phi$ when comparing two models.
In other words, a fair approximation to the Bayes' factor can be obtained by simply comparing the likelihood ratios,
\be
\log\frac{P(D|M_1)}{P(D|M_0)}\approx ({\rm MLR})_1-({\rm MLR})_0.
\ee

The model that best fits the INTEGRAL/SPI is a NFW profile with constant cross-section.
This model closely ressembles the halo model $H'$ of \citet{Knodlseder05}, with similar values of the best-fitting flux and MLR.
All other models in Table~\ref{tabSingle} have $\Delta_\MLR>20$ with respect to the $a$-term NFW or any of the bulge and halo models in \citet{Knodlseder05} that satisfactorily explain the observed morphology of the 511~keV emission.
In frequentist terms, all the other density profiles and dark matter types listed in Table~\ref{tabSingle} can be rejected at the $(1-e^{-20})$ confidence level, or $\sim\sqrt{20}\,\sigma$.

A rigorous Bayesian comparison between the one- and two-parameter models would require an extensive discussion of the priors.
However, we note that linear combinations of the $a$-term and $b$-term models (c.f. Table~\ref{tabCombined}) lead to marginal improvements of the MLR, but they provide non-physical solutions where one of the fluxes is negative.
When the fit is constrained to non-negative values, zero fluxes are obtained for the negative coefficients, and the results quoted in Table~\ref{tabSingle} are reproduced for the other component.
We therefore conclude that two-parameter models fail to provide a better explanation of the observational data.

%--------------------------------------------------------------------------
   \section{Results} \label{secResults}
%--------------------------------------------------------------------------

%____________________________________________
\subsection{The nature of dark matter}

Comparing the best-fitting flux in Table~\ref{tabSingle} for a NFW profile with an $a$-term with our theoretical prediction in Table~\ref{tabfluxes}, we obtain
\be a_{26}=(2.6\pm 0.12)\times10^{-4}\ \mmev^2.
\label{a26spi}
\ee

This result is valid for any kind of candidates, as long as the
annihilation cross-section is parameterized as we have done in the
present work and the galactic positrons are produced as primaries.

Equation~(\ref{a26spi}) becomes comparable with (\ref{eqRelic}) when $\mdm
\simeq$ 100 MeV. This confirms our previous conclusions from the
gamma ray constraints, i.e. that the LDM scenario requires both an $a$
and $b$-term when $\mdm < 100$~MeV.

To explain the 511 keV emission line, candidates heavier than 100
MeV would require a (velocity-independent) annihilation
cross-section into a pair electron-positron that is well above the
relic density requirement.
Heavy candidates must thus produce positrons as secondaries to be a possible solution.
However, one would expect an overproduction of gamma rays in the Milky Way in such case, unless there existed a channel that lead to a large production of positrons and a low production of gamma rays.
This is why we advocate for LDM particles with $\mdm < 100$~MeV in order to explain the origin of galactic positrons.

For a scalar particle, the F-exchange channel yields \be
 \sigma v_r
 = \frac{c}{4 \, \pi} \sqrt{1\!-\!{\left(\!\!\frac{m_{\rm f}}{\mdm}\!\!\right)}^{\!\!2} }
 \left[ 1\!-\!{\left(\!\!\frac{m_{\rm f}}{\mdm}\!\!\right)}^{\!\!2}\!+\!v\dm^2\right]
 \!\frac{c_l^2 c_r^2}{m_F^2}\
\ee where $m_{\rm f}\ll\mdm$ is the mass of the fermions in the
final state, $m_F$ is the mass of the heavy fermion that is
exchanged during the annihilation and the quantities $c_l$ and $c_r$
correspond to the couplings.

Neglecting $m_f$ compared to $\mdm$, we find
\be
\label{a26scalar}
a \simeq c \ \frac{ c_l^2\,c_r^2\ }{ 4 \ \pi\ m_F^2} \Rightarrow
a_{26} \ \simeq \ 9.32\times10^{3} \ c_l^2\,c_r^2\ \left(\frac{m_F}{100 \, \rm{GeV}}\right)^{-2}
\ee

Substituting (\ref{a26scalar}) into (\ref{a26spi}) yields
\be
\label{equalsq}
\frac{m_F}{100\ \rm{GeV}} \simeq 6\times10^{3} \ \frac{c_l\,c_r}{\mmev},
\ee
where the two couplings $c_l$ and $c_r$
are expected to be lower than unity (a few units at most).

For $c_l \approx c_r \approx 1$, one obtains $m_F \approx 6-O(600)$
TeV (for $1\la\mmev\la100$). This is obviously out of reach for past
and forthcoming colliders. However, for smaller but more realistic
values of the couplings, $m_F$ could be within the range of next
colliders. According to equation (\ref{equalsq}), a mass $m_F\in
[100 \mbox{GeV},O(\mbox{TeV})]$ would correspond to $c_l\,c_r\in
1.67 \times [10^{-4}, 10^{-3}] \, \mmev$. Those are not particularly
small couplings (especially for $\mmev \sim 100$~MeV) so there might
be a signature of the LDM scenario in the next generation of
accelerators, and in particular at LHC if the $F$ particles
associated with quarks ($F_q$) are not too heavy and their couplings
not too small. Note that smaller couplings $e-F_e-dm$ (leading to
$m_{F e} < 100$ GeV) are forbidden due to the absence of signal in
past colliders (notably LEP).

So far we have considered scalar LDM particles. However, fermionic
candidates (either Dirac or Majorana) have also been suggested
\citep{bf}. The difference with respect to the scalar case would be
that, for a Dirac fermion, the cross-section would be given by \be
 a_{\rm D} \sim \frac{c}{32\,\pi}\,{(c_l^2 + c_r^2)}^2\ \frac{\mdm^2}{m_F^4}
\ee
instead of (\ref{a26scalar}), while for Majorana particles,
\be
 a_{\rm M} \sim \frac{c}{8\pi}\,(c_l^2 \, c_r^2)\
 \frac{\mdm^2}{m_F^4},
\ee where $F$ now denotes a scalar.

Using equation (\ref{a26spi}), we obtain that the mass $m_F$ should
be \be \frac{m_F}{100\ \rm{GeV}}\sim 0.145 \ \sqrt{(c_l^2 + c_r^2)}
\ee for Dirac dark matter particles and \be \frac{m_F}{100\
\rm{GeV}} \sim 0.206 \ c_l \, c_r \ee for Majorana candidates.

Assuming realistic values for the couplings (i.e. $c_{l, r}<O(1)$),
we find that the $F$ mass is much smaller than 100 GeV regardless of
whether LDM is a Dirac or Majorana fermion. Since the presence of
charged particles much lighter than $\sim 100$~GeV has been excluded
by LEP data, one readily sees that fermionic LDM particles cannot
explain the 511 keV line emission unless one considers couplings at
the edge of perturbativity.

Our results also indicate that a $Z'$ cross-section cannot explain
the observed 511 keV emission on its own. A cross-section strictly
proportional to $v^2$ can be ruled out by $\Delta_\MLR \geq 29.9$.
The best-fitting fluxes obtained for a linear combination of the
models with an $a$- and a $b$-term are given in
Table~\ref{tabCombined}, where two free parameters have been
adjusted to match the SPI data. Although the MLR is somewhat
improved, a non-physical solution (with one of the factors being
negative) is always obtained, suggesting that the boson-exchange
channel plays only a minor role within the Milky Way halo.

The existence of scalar DM coupled to heavy (fermionic) particles is
thus required for our model in order to explain the 511 keV line, at
variance with the results reported by \citet{boehmhooper} where the
$F$ were thought to be facultative, but in agreement with
\citet{boehmyago}. The reason for the discrepancy resides in the
more accurate description of the velocity dispersion profile of the
Milky Way.
%\cb Unfortunately, the negligible contribution of the
%$b$-term to the total cross-section makes it impossible to
%discriminate between our version of the LDM scenario and the variant
%proposed by \citet{komatsu}, which does not involve any $Z'$ atall.\cb

For the density profiles considered in the present work, decaying
dark matter is completely incompatible with the observed morphology
of the 511 keV emission. Even for the best-fitting model, M99,
decaying dark matter of any sort can be ruled out by
$\Delta_\MLR=127.9$. It would be possible, though, that a steeper
dark matter halo (e.g. as predicted from adiabatic contraction) may
provide a better match to the observations \citep[see
e.g.][]{Prada04}.

%____________________________________________
\subsection{The Milky Way density profile}

All of the configurations in Table~\ref{tabSingle} display
$\Delta_\MLR>20$ when compared to our best-fitting model, namely a
NFW halo of annihilating dark matter with approximately constant
cross-section. We will now assume that this is indeed the nature of
dark matter particles.
%\footnote{Note that this is a fairly general
%condition, which would be satisfied not only by the LDM scenario.}.
If most galactic positrons did arise from DM annihilations, the
spatial distribution of the 511 keV emission line would provide an
extremely sensitive probe of the shape of Milky Way dark halo.

\begin{figure}
\centering
\includegraphics[width=8cm]{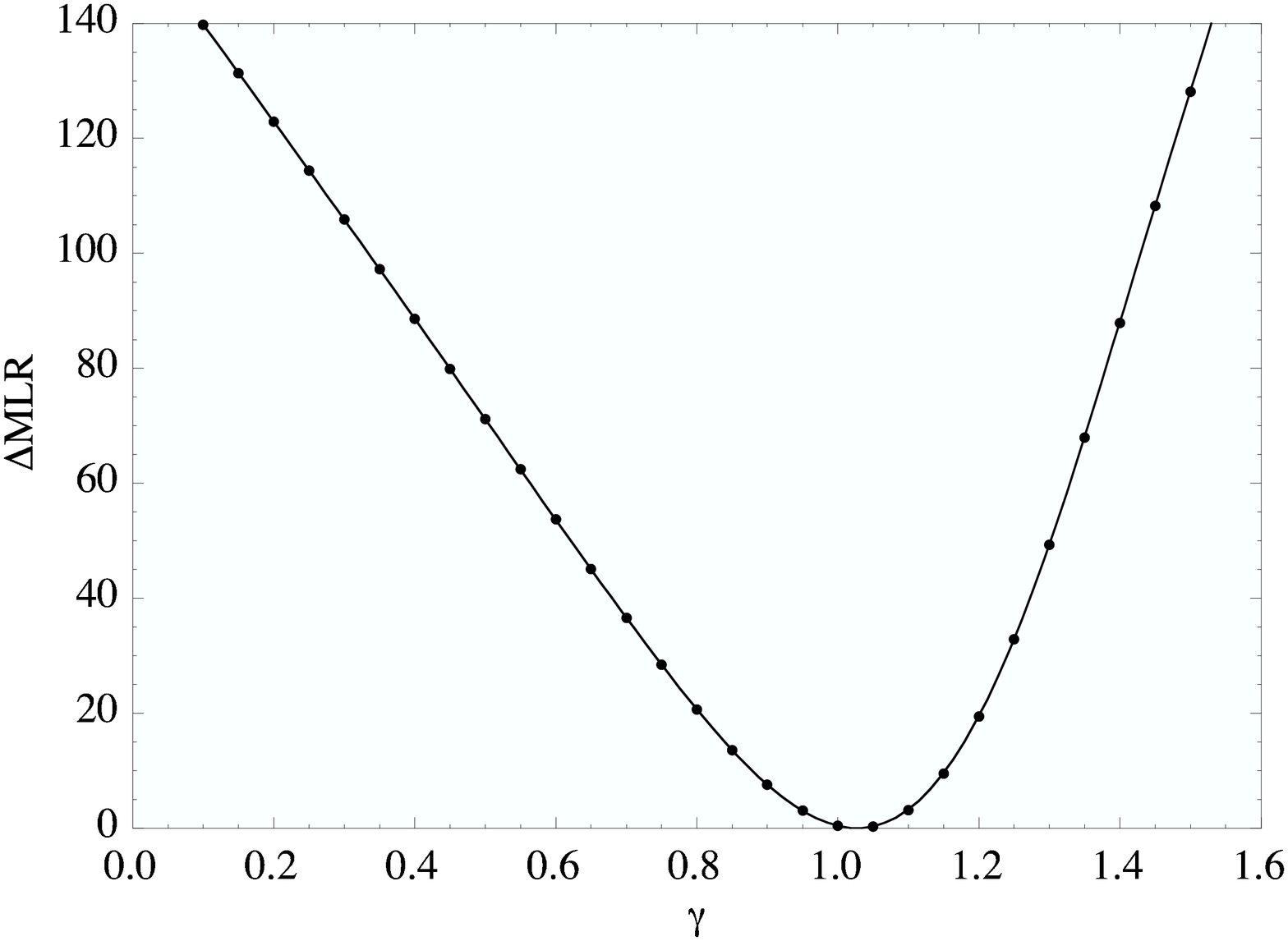}
\caption{ $ \Delta_\MLR=\rm{MLR}(\gamma)-\rm{MLR}(1)$ for different
central asymptotic slopes $\gamma$. Continuous line plots the
best-fitting ninth-order polynomial. } \label{figGamma}
\end{figure}

We plot in Figure~\ref{figGamma} the MLR of a series of models in which the parameters $\alpha$ and $\beta$ in expression (\ref{eqrho}) have been set to $\alpha=1$ and $\beta=3$, whereas the inner slope $\gamma$ varies from $\gamma=0.1$ to $\gamma=1.6$ in uniform steps $\Delta\gamma=0.05$.
We have normalized the MLR to the particular case $\gamma=1$, which corresponds to the NFW profile.
Fitting the data points with a ninth-order polynomial, the optimal value of the logarithmic slope is found to be $\gamma=1.03\pm0.04$, where the errors have been estimated by equating $\Delta_\MLR=1$.

This result is perfectly compatible with $\gamma=1$, but any of the other profiles suggested in the literature would be extremely hard to reconcile with the INTEGRAL/SPI data.
This is again at odds with \citet{boehmhooper}, where a shallower profile with $\gamma\sim0.6$ was favoured, based on a coarser comparison between the theoretical predictions and the observed flux and extension of the emission.

Finally, we would like to stress that the constraint we obtain for
the inner asymptotic slope of the density profile is so tight that,
if the Milky Way dark halo was found to follow a different shape by
some independent means, the possibility that dark matter
annihilations were the main source of galactic positrons would seem
rather unlikely. Systematic effects (see Section~\ref{secSyst})
would in general tend to yield values of $\gamma$ below the real
one, so our estimate should be regarded, to a certain extent, as a
lower limit. If DM is responsible for the 511 keV emission,
$\gamma\ga1$. If $\gamma<1$, galactic positrons must come from a
different physical process.

%--------------------------------------------------------------------------
   \section{Discussion} \label{secDiscus}
%--------------------------------------------------------------------------

%____________________________________________
\subsection{Systematic effects} \label{secSyst}

The morphology of the galactic 511 keV line emission provides a
wealth of information on both the nature of dark matter and the
shape of the Milky Way dark halo. We have shown in the previous
section that several constraints can be derived for the parameters
that characterize dark matter particles (with special emphasis on
the LDM scenario), as well as the inner logarithmic slope of the
radial density profile.

One should keep in mind that our analysis is based on several
simplifying assumptions. Relaxing one (or each) of them would have a
different systematic effect on our results:

First, we have neglected any astrophysical contribution to the galactic positron budget.
Emission from sources other than dark matter would lower our estimate of $a_{26}\,\mmev^{-2}$, equation (\ref{a26spi}), by a factor proportional to the fraction of DM-related positrons.
The effect on the density profile derived for the Milky Way dark halo depends on the spatial distribution of the other sources.
Our estimate of the inner slope $\gamma$ would be biased low if the latter was flat, and high if the sources were concentrated near the centre.

Second, the fact that positrons may travel a certain distance before
losing their energy and annihilating would flatten the expected
emission. Therefore, a steeper density profile would be required in
order to fit the observations (i.e. our estimate of $\gamma$ would
be biased low). The best-fitting flux would be somewhat lower, so
the real value of $a_{26}\,\mmev^{-2}$ would again be lower than
(\ref{a26spi}).

Third, our model of the Milky Way is overly simplistic in several respects.
On one hand, it is well known that the dark matter haloes found in numerical simulations display a significant degree of triaxiality \citep{JingSuto02}, although the inclusion of gas cooling tends to yield more spherical haloes \citep{Kazantzidis04}.
In our galaxy, observations of the Sagittarius tidal stream have been interpreted as favouring a spherical halo \citep{Ibata01,Majewski03}, but recent analyses also suggest both oblate \citep{Martinez04} and prolate \citep{Helmi04} shapes.
Although the INTEGRAL/SPI data is consistent with spherical symmetry, higher-resolution observations would be needed in order to quantitatively address this issue.

 In addition, some substructure is expected to be present in the
dark matter halo of our galaxy, both in real and phase space (i.e.
the six-dimensional space of positions \emph{and} velocities). In
real space, dark matter clumps would tend to boost the expected
 emission due to the increase in local density \citep{Bergstrom99}.
It has been argued \citep[e.g.][]{Hooper04} that the 511 keV line
could actually be detected not only from the galactic centre, but
also from the nearest dwarf spheroidals. On the other hand,
structures in phase space (such as tidal streams) would not have any
effect on the emission through the $a$-channel, but they may have a
significant impact both on direct detection experiments
\citep{Helmi02} and on the emission arising from the $b$-term of the
cross-section, mostly because of the increase in dark matter
velocity with respect to the local velocity dispersion. Although it
does not seem likely, for $b_{26} \sim 10$, that structures in
velocity space yield a detectable signature, there might be a mild
enhancement in the signal from local dwarfs, particularly near their
pericentre.

A more realistic model of the density and velocity distribution of dark
matter particles within the Milky Way halo would be given by the
results of N-body simulations. Although extremely promising, this
approach \citep[see e.g.][]{Stoehr03} must face the problem of
numerical convergence (i.e. lack of resolution) in order to provide
conclusive results.

%____________________________________________
\subsection{Compatibiliy with low-energy gamma rays} \label{secGamma}

DM annihilations produce charged particles which, in turn, produce
low-energy gammas. But low-energy photons are also expected through
the final state radiation (FSR) mechanism or DM annihilations into
two monoenergetic photons (with an energy $E=\mdm c^2$). In principle,
the observation of such line would be an unambiguous signature of the LDM scenario.
However, the corresponding cross-section relies on a box diagram that involves four powers of the electric charge $e$, and therefore it is expected to be suppressed compared to the FSR mechanism in which either the electron, the
positron or the charged particle that is exchanged during the
annihilation emits a photon.

A rough estimate of the gamma ray flux generated by FSR was given in
\citet{bens}. In an attempt to build a model that would
surely satisfy astrophysical constraints, the authors assumed that
the FSR cross-section was as large as the annihilation cross-section.
The electric charge (which is expected to reduce the FSR
cross-section by a factor $e^2$) was therefore deliberately omitted in order to obtain the
maximal flux of low-energy gamma rays that LDM could produce.
Comparing this estimate with observations, \citet{bens} deduced that
particles lighter than 100 MeV were excluded unless their
annihilation cross-section was suppressed in our galaxy with respect to
its value in the primordial universe, which led the authors
to postulate the existence of a velocity-dependent annihilation
cross-section.

The required suppression factor turns out to be consistent with
that obtained from the analysis of the 511 keV line. More precisely, a
cross-section as large as equation (\ref{a26spi}) does not overproduce
gamma rays if the DM particles are lighter than 30 MeV. It does, in
principle, if DM is heavier, but given the drastic assumption on
the FSR cross-section, it was considered that even 100 MeV particles
would both explain the 511 keV line and be compatible with gamma-ray
observations.

In a recent paper, \citet{Beacom05} claimed that particles
heavier than 20 MeV were actually not consistent with the gamma-ray data.
To make this point, they wrote the FSR cross-section as
\be
\frac{\dd\sigma_{\rm b}}{\dd E_\gamma} = \sigma_{\rm ann} \times R_{\rm corr}(E)
\label{eqFSR}
\ee
and used $R_{\rm corr}(E) =
~\frac{\alpha}{\pi} ~\frac{1+{(s'/s)}^2}{E_\gamma}
 \left[ \ln\!\left(\!\frac{s'}{m_e^2}\!\right)\!-1 \right]$,
with $s\equiv4\mdm$ and $s'\equiv4\mdm(\mdm-E_\gamma)$,
as computed by \citet{berends} for the radiative correction to
$e^+ e^- \rightarrow \mu^+ \mu^-$.

To our knowledge, it has never been proven that any radiative
cross-section can be written in the same way as $e^+ e^- \rightarrow
\mu^+ \mu^- \gamma$ in the hard photon limit, so
it might be possible that equation (\ref{eqFSR}) does not provide a completely accurate description of the radiative corrections to the annihilation process.
Nonetheless, a detailed computation would be quite delicate, and it certainly lies well
beyond the scope of the present work.
We will therefore estimate the gamma ray flux according to the procedure followed by \citet{Beacom05}.
Assuming the same $R_{\rm corr}(E)$, the intensity of internal bremsstrahlung gamma rays at the earth is related to the 511 keV skymaps by
\be
 \frac{\dd I_{\rm b}}{\dd E_\gamma} = \frac{I_{511}}{0.605} ~\frac{\alpha}{\pi}
~\frac{1+{(s'/s)}^2}{E_\gamma}
 \left[ \ln\!\left(\!\frac{s'}{m_e^2}\!\right)\!-1 \right].
\label{eqGam0}
\ee

Comparison with COMPTEL and EGRET data only requires the computation of the average intensity over the appropriate region of the sky.
Using the model that best fits the 511 keV line, i.e. a NFW profile with constant cross-section given by equation~(\ref{a26spi}), we obtain
\be
\langle I_{511}\rangle_{|l|<30\degr,|b|<5\degr}=
\frac{\Phi_{511}}{\Delta\Omega}\approx 0.005~{\rm cm^{-2}~s^{-1}~sr^{-1}}.
\label{eqGam1}
\ee
Even if we restrict to the innermost $5\degr$ in both longitude and latitude, our expectation for the average intensity
\be
\langle I_{511}\rangle_{|l|<5\degr,|b|<5\degr}
\approx 0.018~{\rm cm^{-2}~s^{-1}~sr^{-1}}
\label{eqGam2}
\ee
is somewhat lower than the estimate used by \citet{Beacom05}.

\begin{figure}
\centering
\includegraphics[width=8cm]{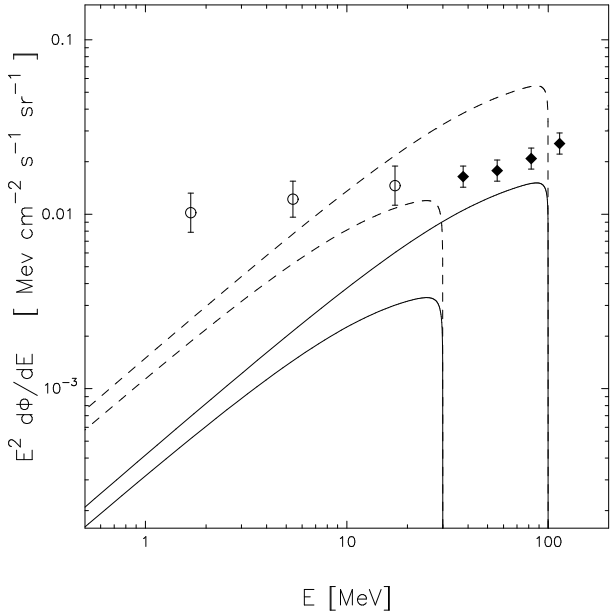}
\caption{
Comparison with COMPTEL (circles) and EGRET (diamonds) data \citep{Strong00}.
Lines plot the gamma ray flux expected for $m\dm=30$ and 100~MeV, according to expression~\ref{eqGam0}.
Solid and dashed lines correspond to the regions $|b|<5\degr$, $|l|<30\degr$ (\ref{eqGam1}) and $|b|<5\degr$, $|l|<5\degr$ (\ref{eqGam2}).
}
\label{figGammaRays}
\end{figure}

The solid angle considered and the precise shape of the energy
dependence of the observational data have important consequences on
the inferred constraints. Figure~\ref{figGammaRays} shows the limits
we obtain for the mass of the dark matter candidate. The factor of $\sim4$
difference between equations~(\ref{eqGam1}) and~(\ref{eqGam2})
translates into a factor of 3 for the maximum dark matter mass, i.e.
$\mdm\la[30-100]$~MeV. This constraint is similar (though less
severe) to that obtained by \citet{Beacom05}, and it also agrees
roughly with the results reported by \citet{bens} despite the fact
that much less than 2 photons are predicted by integrating
equation~(\ref{eqGam0}) over the photon energy (e.g. between 1 and 100 MeV).
However, the constrain provided by the `spectrum' of the observed gamma-ray radiation is much more stringent than its total flux \citep[or, more precisely, the flux above $E_\gamma^{\rm min}\le\mdm c^2$, see][]{bens}.

In any case, we would also like to emphasize that a rigorous comparison with COMPTEL and EGRET data,
analogous to the analysis of the 511 keV line presented here, would
be necessary in order to provide accurate constrains on the mass
range allowed for the dark matter particles.

%____________________________________________
\subsection{Tests of the LDM model in particle physics experiments}

We mentioned in a previous section that the $F$ particles could be
detectable in the next colliders. The $F$ particles could be
detected through their two-body decay ${F_{p}}^- \rightarrow p^- +
\slash{\hspace{-0.2cm} E}$, where $\slash{\hspace{-0.2cm} E}$
denotes the missing energy associated with Dark Matter and $p^-$ the
Standard Model particle associated with $F_{p}$ (for example
${F_{e}}^- \rightarrow e^- + \slash{\hspace{-0.2cm} E}$). The
couplings are large enough to allow for the decay within the
detector, although the production cross-section of $F$ particles
($p^+ p^- \rightarrow F_{p}^- F_{p}^+$) may be too small to yield a
visible signature. This cross-section can reach, however, a few pb
for $m_{F p} \sim 100$ GeV, $\mdm\sim 100$ MeV and couplings
$F_p-p-dm$ of the order of (\ref{equalsq}). The presence of $F_q$
particles could then be detected at LHC through the modification of
the total hadronic cross-section (if either of the couplings $c_l$
or $c_r$ remains relatively large).

The $F_e$ decay should be quite similar to the (supersymmetric)
decay of a chargino into an electron and a sneutrino. However, in
our case the mass of the DM particle is much smaller than the
sneutrino mass. Hence, we expect a fair repartition of energy
between the electron and the missing energy in the LDM scenario.
This should be quite different in a supersymmetric framework, albeit
more precise estimates would depend on the difference of mass
between the sneutrino and the chargino (and therefore on the model
considered). In any case, a possible way to discriminate between our
scenario and supersymmetric particles could be the absence of other
signatures. For example, unlike the chargino which is expected to
also decay into a selectron and a neutrino, the $F_e$ particles are
assumed to leave only one main signature (i.e. the decay into an
electron and missing energy). The precise relationship between the
$F_e$ mass and the couplings (\ref{equalsq}) should also help in
discriminating between the LDM and supersymmetric scenarios.

Dark matter production through $p^+ p^- \rightarrow {\rm dm\,dm}$
could in principle be detectable in the initial state radiation
process (where a photon is emitted by the particles of the initial
state). However, the associated cross-section is expected to be
smaller than a few fb even for $\mdm \sim$ 100 MeV, and is therefore
less interesting.

There are two other tests that could be really crucial for our
purpose. One is based on the so-called NuTeV anomaly and the other
one on the value of the fine structure constant $\alpha$.

NuTeV is an experiment which measured the ratio $$R=
\frac{\mbox{neutral currents}}{\mbox{charged currents}} = (g_l^2 -
g_r^2),$$ with $g_{l,r}^2= [(g_{l,r}^u)^2+ (g_{l,r}^d)^2]/4$ and
$g_{l,r}^f = 2(T_3(f_{l,r}) - Q(f) \sw^2)$ the left and right
couplings of the $Z$ boson to fermions. Through the measurement of
this ratio, one can infer the value of the mixing angle $\sw^2$ and
compare it to Standard Model expectations (obtained notably from LEP
 electroweak precision observables).  Surprisingly, NuTeV result
for $\sw^2$ turns out to be slightly discrepant with the Standard
Model value \citep{nutev1,nutev2,nutev3}.

There are several possible explanations to the NuTeV anomaly. In
particular, isospin violation and/or strange sea asymmetry as well
as other effects such as electroweak corrections may reduce
significantly the discrepancy. However, at present, the situation is
still uncertain and there is an open window for new physics. The
best explanation
 to that respect turns out to be a light gauge boson \citep{sacha,boehm}, similar to the one we introduce in
 the LDM scenario. If the NuTeV anomaly disappears or is significantly
 reduced, then one could set a limit on the $Z'$ couplings to ordinary
 matter. On the other hand, more evidence in favour of a light $Z'$ would
certainly give more credit to the LDM scenario.

The other test concerns the existence of heavy particles
\citep{boehmyago}. We expect them to contribute to the electron
anomalous magnetic moment as: $\delta a_{\mu, e} \sim \frac{c_l
c_r}{16 \pi^2} \frac{m_{\mu, e}}{m_{F_{\mu, e}}} >0$.  Using
expression~(\ref{equalsq}), we obtain
\begin{equation}
\label{dath} \delta a_{e}^{F} \sim 5.41\times10^{-12}\ \frac{\mdm}{\rm{MeV}}.
\end{equation}

It turns out that there is a small discrepancy between the
theoretical value of $a_e$ (hereafter denotes $a_{th}$) and its
measurement ($a_{exp}$):
\begin{equation}
\label{daexp} \Delta a_e = (a_{exp} - a_{th}) \sim (3.44-3.49) \
10^{-11}
\end{equation}
where the first number is obtained from the positron g-2, while the
second one is from the electron. We estimate (\ref{daexp}) by using
$a_{th} = f(\alpha)$ with $\alpha = \alpha_{QH}$, the fine structure
constant as measured by Quantum Hall effect (QH) experiments. There
are other experiments aiming at measuring $\alpha$, but QH
experiments seem the most precise at present \citep[see e.g.][]{kinoshita}.

Usually, one assumes instead the validity of QED and imposes that
$a_{th} \equiv a_{QED}$ matches $a_{exp}$. One then gets a
`theoretical' estimate of $\alpha$ that is extremely precise and
in fact the most significant input in the $\alpha$ value given in
the CODATA. The latter differs too much from $\alpha_{QH}$ for the
difference to be explained by common extensions of the Standard
Model such as supersymmetry, but experiments measuring $\alpha$ did
not reach the sensitivity of $g-2$ experiments as yet. Therefore,
there is the hope that the difference may eventually go away and
that there is no new contribution other than QED.

%However, in the absence of physical
%explanations, it is sometimes argued that this discrepancy will
%probably go away when experiments measuring $\alpha$ will reach the
%sensitivity of $g-2$ experiments.
%This may be true.

However, the discrepancy between $\alpha_{QH}$ and $\alpha$ could
originate from new physics. In particular, the introduction of heavy
particles coupled to light scalars adds a new contribution in
$a_{th}$ which is greater than expected in e.g. supersymmetry, due
to the DM mass scale.

Including this new contribution ($\delta a_{e}^{F}$),  one obtains
$\alpha \simeq \alpha_{QH}$ (or $a_{th} \simeq a_{exp}$)
if $\mdm\sim6.4$ MeV, using the average value of $a_{th}$ computed with
$\alpha_{QH}$ \citep[see][]{kinoshita}.
Taking into account theoretical and experimental uncertainties, $\mdm\sim3-9$~MeV.
For smaller DM masses, we
obtain $\delta a_{e}^{F} \leq \Delta a_{e}$, while for larger DM
masses $\delta a_{e}^{F} \geq \Delta a_{e}$. As explained
above, there is no direct measurement of $\alpha$ that is as precise
as the $g-2$ as yet. Therefore, it is hard to exclude values above 7
MeV. However, this certainly places a very strong contraint and
motivates further experiments measuring the value of the fine
structure constant directly (and independently of QED). If these
experiments find a perfect agreement with the value recommended in
the CODATA, then the LDM scenario will have difficulties in
explaining the 511 keV line emission. If they found a discrepancy
(whether it is positive or negative) then LDM will remain a serious
candidate because it would be the sign of new physics. In
particular, if the value $\alpha_{QH}$ is confirmed, then the LDM
scenario may reconcile the results from both $g-2$ and $\alpha$
experiments, despite the difference of sensitivity.

%Higher-order recoil corrections to helium fine structure
Measurements of the fine structure interval of $H_e$-like ions based
on laser spectroscopy might also provide a very precise
determination of $\alpha$ if theory gets as accurate as the
experimental determinations \citep[see e.g.][and references therein]{pachucki}.
However, this method assumes that there is no
additional contribution from new physics, so it may not be suitable
for answering whether the $F$ particles exist or not.

Taking the same couplings and the same mass $m_F$ for $F_e$ as for
$F_{\mu}$, we obtain a very large contribution to the muon $g-2$.
Our prediction, in fact, exceeds the experimental value by a factor
2-3, which is itself larger than the Standard Model prediction
\citep{mugm2}. It was found $\Delta a_{\mu} = (a_{exp} -a_{th}) \in
[1.6, 2.7] \ 10^{-9}$. So, by using $\mmev \sim 6-7$ and $ m_{F \mu}
= 3 m_{F e}$ (or e.g. $ m_{F \mu} = 2 m_{F e}$ and smaller couplings
to the muons), our prediction for the muon $g-2$ becomes compatible
with the experimental value. In fact, the LDM scenario would even
explain the well-known discrepancy. Note that such a hierarchy
exists in the Standard Model and it is very realistic to assume that
it exists also in any other extensions.

Hence, the LDM scenario could in fact explain both the experimental
values of the fine structure constant and the muon $g-2$ for $\mdm\approx 6-7$ MeV.

%--------------------------------------------------------------------------
  \section{Conclusions} \label{secConclus}
%--------------------------------------------------------------------------

In this paper, we use the intensity and morphology of the observed
511 keV line to put independent constraints on the nature of LDM and
the shape of the Milky Way dark halo. Our main assumptions are that
DM decays or annihilations are the main source of galactic positrons
and that these positrons do not travel long distances before annihilating.
Theoretical expectations for the flux distribution are
computed for different DM models and galactic density profiles. The
sky maps thus obtained are then convolved with the SPI response
function and used as a source for the INTEGRAL model-fitting
analysis.

We can rule out from a likelihood analysis the possibility that
decaying dark matter is responsible for the observed emission,
unless the density profile of the Milky Way dark halo turns out to
be extremely cuspy (with inner asymptotic slope $\gamma>1.5$).

We can exclude fermionic LDM particles, because it would require the
introduction of charged scalars lighter than 100 GeV, which should
have already been detected in past colliders. As a result, LDM is
likely to be a scalar.

For annihilating scalar LDM, it is shown that the exchange of a
heavy fermion ($F_e$) is required in order to fit the morphology of
the 511 keV line, while the existence of a $Z'$ boson would be
necessary to satisfy the relic density criterion. Assuming a full
spectrum and, most precisely, the existence of $F$ particles
associated with quarks ($F_q$), we notice that there might be a
signature at the Large Hadron Collider, notably through the measure
of the total hadronic cross-section and the two body decay of these
$F_q$ particles.

However, the most promising signature of $F$ particles turns out to
be their contribution to the electron $g-2$. The new contribution
would make the measurements of the fine structure by the Quantum
Hall experiment and the electron anomalous magnetic moment
compatible for $\mdm\sim 6-7$~MeV, meaning that the value of
$\alpha$ quoted in the CODATA (and used for many estimates) may not
be the correct one.
The dark matter mass could however be larger than $\sim$ 6 MeV.
The existence of clumps, the fact that the dark halo is probably not perfectly spherical and the contribution to the positron population from astrophysical sources are all expected to decrease our estimate in equation~\ref{a26spi}, therefore allowing for larger dark matter masses.

Assuming the existence of this spectrum (and $F_{\mu}$ particles),
we also find a non-negligible contribution to the muon $g-2$. Both
$F_e$ and $F_{\mu}$  could then explain the discrepancy between the
Standard Model predictions and the experimental values of the muon
$g-2$ and the fine structure constant. Alternatively those could
provide a way to constrain the LDM scenario.

Concerning the shape of the Milky Way dark halo, our results clearly
indicate that dark matter particles can only explain the observed
511 keV emission if our galaxy features a cuspy density profile. For
any annihilating DM candidate with constant cross-section, the
best-fitting inner asymptotic slope is found to be
$\gamma=1.03\pm0.04$.

To sum up, we would like to stress the fact that the 511 keV
emission line provides extremely stringent constraints on the light
dark matter parameters. Independent confirmations are needed to
prove that dark matter contributes to most of the galactic
positrons. Such confirmation might come either from the lack of
astrophysical sources and/or from detection in particle physics
experiments. Observations of the density profile of the Milky Way
have the possibility to rule out a dark-matter related origin of
galactic positrons if the density profile of our galaxy is found to
be shallow at the centre. Alternatively, the discovery of LDM
particles would have a tremendous impact on the determination of the
dark halo profile of the Milky Way.

%---------------------------------------------------------------

\section*{Acknowledgments}

The authors would like to thank J. Devriendt, F. Ferrer, P. Janot and P.
Salati for useful discussions. Y. Ascasibar acknowledges support
from NASA grants G02-3164X and G04-5152X.

%--------------------------------------------------------------------------
 \bibliographystyle{mn2e}
 \bibliography{511keV}

\end{document}